\documentclass{emulateapj}
\shorttitle{}
\shortauthors{Wu et al.}
\begin{document}
\title{Orbits of nearby planetary nebulae and their interaction with the interstellar medium}
\author{Zhen-Yu Wu, Jun Ma, Xu Zhou}
\affil{Key Laboratory of Optical Astronomy, National Astronomical Observatories, Chinese Academy of Sciences, 20A Datun Road, Beijing 100012, China}
\and
\author{Cui-Hua Du}
\affil{College of Physical Sciences, Graduate University of the Chinese Academy of Sciences, Beijing 100049, China}
\email{zywu@nao.cas.cn}
\begin{abstract}
We present and analyze the orbits of eight nearby planetary nebulae (PNs) using two different Galactic models. The errors of the derived orbital parameters are determined with a Monte Carlo method. Based on the derived orbital parameters, we find that Sh 2-216, DeHt 5, NGC 7293, A21, and Ton 320 belong to the thin-disk population, and PG 1034+001 and A31 belong to the thick-disk population. PuWe 1 probably belongs to the thick-disk population, but its population classification is very uncertain due to the large errors of its derived orbital parameters. The PN-ISM interactions are observed for the eight PNs in our sample. The position angles of the proper motions of the PNs are consistent with the directions of the PN-ISM interaction regions. The kinematic ages of PNs are much smaller than the time for them to cross the Galactic plane. Using the models of Borkowski et al. and Soker et al., the PN-ISM interaction can be used to derive the local density of ISM in the vicinity of evolved PNs. According to the three-dimensional hydrodynamic simulations of  Wareing et al. (WZO), Sh 2-216, A21, and Ton 320 are in the WZO 3 stage, PG 1034+001 and NGC 7293 are in the WZO 1 stage, and PuWe 1 is in the WZO 2 stage.
\end{abstract}

\section{Introduction}
Planetary nebulae (PNs) are the descendants of asymptotic giant branch (AGB) stars. According to the Interacting Stellar Wind (ISW) model \citep{kw78,kw82}, an AGB star loses most of its envelope mass with a slow stellar wind (AGB wind) until the hot core is exposed. After reaching the PN nucleus stage, a new faster wind (post-AGB wind) sweeps up the circumstellar material originating in the slow wind and forms the high-density and ionized nebular shell. Most PNs are not spherical and exhibit complex shapes: round, elliptical, butterfly, bilobed, and irregular. Stellar rotation, magnetic field, and binarity are important ingredients in shaping PNs \citep{de09, ba02}.

The interaction of a moving PN with the interstellar medium (ISM) was first discussed by \citet{gu69}. \citet[hereafter BSS]{bo90} examined the morphology of nine PNs with a center star (CS) whose proper motions exceeded 0.015 arcsec yr$^{-1}$ and found that seven of the examined PNs interact with the ISM. They also examined the morphology of large, nearby PNs, and found that many were asymmetric. \citet{tw96} presented a CCD imaging atlas of large PNs ($>8'$) and found that 21 of 27 PNs in their sample revealed significant interaction with the ISM. Using deep narrowband CCD images of the nine largest, nearby PNs, \citet{xi96} found that eight PNs in their sample revealed features indicating an interaction between the PNs and the ISM. More recently, \citet{ra09} presented mid-infrared images for five PNs which appear to interact with the ISM. Using the data of the INT Photometric H$\alpha$ Survey (IPHAS) \citep{dr05}, \citet{sa09} identified 21 objects as possible PNs interacting with the ISM.

On the theoretical side, the first theoretical study of the interaction between PNs and ISM was made by \citet{sm76}. Using a thin-shell approximation and simple analytic formulae, \citet{sm76} found that the motion of a PN through the ISM would be expected to cause a shift of the CS toward the leading surface of the interacting PN and ISM. BSS used simple analytical estimates to study the interaction between a moving PN and the ISM and found that the interaction produces asymmetries in the surface brightness of the PN. Using a two-dimensional numerical hydrodynamic code, \citet[hereafter SBS]{so91} simulated the interaction between the PN and the ISM. They considered two limiting cases: adiabatic and isothermal models. In their adiabatic model, a PN moves with large relative velocity ($>100$ km s$^{-1}$) within the ambient ISM with low density so that the cooling time is larger than the flow time. The adiabatic model is relevant for PNs moving in the Galactic halo. In the isothermal model, which is relevant for PNs in the Galactic disk, the PN moves through the high-density ISM with moderate relative velocity $\sim 60$ km s$^{-1}$ and the cooling is very effective. The isothermal model of SBS validates the thin-shell approximation originally developed by \citet{sm76}. SBS also found that the simple analytic model breaks down in their adiabatic model in which a Rayleigh-Taylor (RT) instability develops, leading to shell fragmentation, while the isothermal model shows no sign of the RT instability. \citet{so97} studied the interaction of a PN with a magnetized ISM and showed that the ISM magnetic field, which was not incorporated in the numerical simulations of SBS, can lead to the development of RT instability even in the isothermal environment.

According to the simple thin-shell model for the PNs moving through the ISM, the interaction between PNs and ISM can be divided into three phases \citep{bo90,so91}. First, when the density of the young PN is much larger than that of the ambient ISM, the PN will expand freely. Second, when the PN density drops below a critical value, the ISM pressure becomes comparable to the one in the PN shell and the upstream side of the PN shell will be compressed and the surface brightness in the interaction region will increase. When the density in the PN shell has dropped to a lower value, the shape of the PN shell will be significantly distorted and the CS of the PN will move close to the upstream edge. Finally, the CS of the PN will move outside of the PN shell, and the PN shell will be stripped and mixed with the ISM.

In the PN-ISM interaction models of BSS and SBS, they have assumed that the PN shell has formed before the interaction with the ISM and the interaction only be observable during the late stage of the PN evolution. Moreover, high ISM density, large relative velocity, or both explain the observed asymmetry in the external shell of a PN. \citet{vi03} performed two-dimensional numerical simulations of the evolution of an AGB star moving through its surrounding ISM. They found that during the early stage of the evolution of an AGB star, the slow AGB wind can interact with the ISM. The PN forms later when the fast post-AGB wind sweeps up the stellar material in the slow AGB wind. A spherical innermost shell shaped by the fast post-AGB wind and an external shell with a bow shock character can form. In simulations \citet{vi03} adopted two low ISM densities and a very conservative relative velocity of the PN, 20 km s$^{-1}$. Thus, high ISM density, large relative velocity, and a magnetic field are not necessary to explain the observed asymmetries in the external shells of PNs. Moreover, interaction can occur even during early AGB evolution, which is in agreement with the observed common asymmetries relating to the interaction with ISM in most of the halos of PNs, not just in those of old PNs.

Considering a wide range of PN velocities, mass-loss rates of the CS and ISM densities, \citet[hereafter WZO]{wa07} performed  three-dimensional hydrodynamic simulations, from the beginning of the slow AGB wind phase to the end of the post-AGB/PN phase. They developed a ``triple-wind'' model including a slow AGB wind, a fast post-AGB wind, and a third wind reflecting the movement through the ISM. WZO found that the ISM interaction strongly affects the outer structures of PNs. The PN-ISM interaction process can be divided into four stages (WZO 1 -- 4). In the first stage (WZO 1), the PN expands freely within the bubble of the undisturbed slow AGB wind material and is unaffected by the AGB wind-ISM interaction. In the second stage (WZO 2), the PNs interact with the bow shock formed during the AGB wind-ISM interaction process, and the density and the emission in the upstream side increase accordingly. The third stage (WZO 3) is defined by the geometric center of the PN moving downstream away from the CS. In the fourth and final stage (WZO 4), the PN is completely disrupted and the CS is outside the PN.

The relative velocity of the PN and the ISM density are important parameters for us to consider in order to model and understand the PN-ISM interaction processes. Both of these parameters are relevant to the Galactic orbits of PNs. The orbital parameters can also be used to distinguish different populations in the Galaxy \citep{wu09}, but the available orbital parameters for PNs in the Galaxy are very rare. \citet{ra04} calculated the orbit of PG $1034+001$ and found that this PN is in an advanced stage of interaction with the ISM. \citet{ke04} calculated the orbits of four PNs and confirmed that the motion of the PN shell through the ISM is the origin of the PN-ISM interaction process. In this research, using improved observation data, we calculated the Galactic orbits for eight nearby PNs, which doubles the number of PNs whose orbits are available. Using the derived orbital parameters of those PNs, we also discuss the PN-ISM interaction processes in our sample in detail.

We present the derived Galactic orbits of our PN sample in Section 2. In Section 3, for each PN in our sample, using the derived kinematical data, we discuss the interaction between the PN and the ISM. Our conclusions are given in Section 4.

\section{Galactic orbits}
\subsection{The sample}
The position, distance, absolute proper motion, and radial velocity are essential data for deriving the Galactic orbit of a PN. First, we searched the literature to find PNs whose distance data are available. There are many indirect methods of PN distance determination. On the one hand, statistical distances can be underestimates. On the other hand, spectroscopic distances can be overestimates \citep{be09}. We consider only PNs whose distances are derived from direct parallax measurements. \citet{ha07} presented trigonometric parallaxes of the CSs of 16 nearby PNs. The median error in their derived parallaxes is 0.42 mas which is smaller than the $\pm1$ mas typical error for most \textit{Hipparcos} parallaxes \citep{es97}. Using the \textit{Hubble Space Telescope}, the absolute parallaxes of the CSs of four PNs in the sample of \citet{ha07} were also derived by \citet{be09}. The derived errors in \citet{be09} are 2 -- 3 times smaller than those in \citet{ha07}.

Based on a large number of multi-wavelength data, \citet{fr09} found that the nebula around the CS PHL 932 is a small \ion{H}{2} region and not a real PN.  Rejecting PHL 932 from the sample of \citet{ha07}, we searched the proper motion and radial velocity data for the other 15 PNs in the literature. Using available all-sky astrometric catalogs, \citet{ke08} collected reliable proper motion information (with significance better than $3\sigma$ in at least one component) for a sample of 234 PNs. Five PNs in the sample of \citet{ha07} are not listed in the catalog of \citet{ke08}: NGC 6853, RE 1738+665, PG 1034+001, A7, and A24. We used the VizieR catalog access tool to search the proper motion data for these five PNs. No proper motion data are available for NGC 6853, RE 1738+665, and A24. We found a proper motion for A7 in the UCAC 2 catalog \citep{za04}, but the errors are three times larger than the proper motions. We found a proper motion for PG 1034+001 in the UCAC 3 catalog \citep{za09}, but the proper motions are three times larger than their errors in both components. So, PG 1034+001 is included in the following analysis.

For the 10 PNs included in the catalog of \citet{ke08}, proper motions derived from different astrometric catalogs are listed. Those catalogs include   \textit{Hipparcos} \citep{es97}, \textit{Tycho 2} \citep{ho00}, UCAC 2 \citep{za04}, GSC 2 \citep{la08}, and USNO-B \citep{mo03}. The former four catalogs are all based on the International Celestial Reference System (ICRS), but the proper motions of USNO-B were derived relative to the YS4.0 catalog and are not absolute. We do not adopt the data taken from the USNO-B catalog.

The UCAC 3 catalog was made available in 2009. In addition to the early epoch data used for the UCAC 2 proper motions, UCAC 3 utilizes data from about 5000 unpublished astrographic plates as well as a complete re-reduction of the SPM first epoch data to significantly lower the systematic errors of UCAC 3 proper motions \citep{za09}. The proper motions in UCAC 3 are improved with respect to those in \textit{Hipparcos}, \textit{Tycho 2}, and UCAC 2. Thus, for CS with $V < 14.0$ mag in the sample of \citet{ha07}, the proper motions from UCAC 3 are adopted. For stars with $V > 14.0$ mag, there are larger system errors in the proper motions of UCAC 3. So, for CS with $V > 14.0$ mag in the sample of \citet{ha07}, the proper motions derived from the GSC 2 catalog and listed in \citet{ke08} are adopted, but, the proper motions of NGC 6270 derived from UCAC 2 are very different from those derived from GSC 2, and no data can be found in the UCAC 3 catalog for this object. We rejected this PN in our analysis.

The SIMBAD database was used to search the radial velocity data for the 10 PNs with known distances and proper motions. In the end, eight PNs with reliable distances, proper motions, and radial velocities were included in our final sample. In Table \ref{tb1} the positions ($\alpha$, $\delta$) of each PN adopted from \citet{ke03} are listed in columns 2 and 3, the $V$ magnitudes of the CSs adopted from \citet{ha07} are listed in column 4, column 5 lists the derived parallaxes from \citet{ha07} and \citet{be09}, proper motions ($\mu_{\alpha}\cos\delta$, $\mu_{\delta}$) taken from the UCAC 3 and GSC 2 catalogs are listed in columns 6 and 7, the position angles (P.A.s) of proper motions are listed in column 8, and radial velocities and their references are listed in columns 9 and 10.

In Table \ref{tb21} the angular diameters of the eight PNs, except for PG 1034+001, are taken from \citet{tw96} and are listed in column 2. Using the distances listed in column 4, the linear diameters of these PNs were derived and are listed in column 3. The electron densities, $n_{\textrm{PN}}$, of PNs, if available, are listed in column 5. The volume densities, $n_{\textrm{ISM}}$, of the diffuse warm neutral medium (WNM) and warm ionized medium (WIM) at the observed positions of each PN were calculated according to the model of \citet{ka98} and are listed in columns 6 and 7. The expansion velocities of each PN, if available are listed in column 8, and the corresponding references are listed in the final column. Using the linear diameter and expansion velocity, the kinematic age for each PN was calculated and is listed in column 9. \citet{ka98} used a hydrostatic Galactic model, which assume that the interstellar magnetic field, gas, and cosmic rays form a dynamic equilibrium with the Galactic gravitational potential. They considered various Galactic gas phases in their model; these gases include cold neutral medium (CNM), WNM, WIM, neutral halo medium (NHM), and hot halo medium (HHM). Because the probability of a PN-ISM interaction occurring for CNM is very small \citep{bo90}, and all of the PNs in our sample belong to the Galactic disk population (see the next section), where the densities of NHM and HHM are much lower, these two gases need not be considered in the PN-ISM interaction for disk PNs. Hence we calculated the density for WNM and WIM only with the model of \citet{ka98}. Table \ref{tb21} indicates that the total densities, $n_{\textrm{ISM}}$, of WNM and WIM around the eight PNs are close to $\sim$ 0.1 cm$^{-3}$.

\subsection{Orbital parameters}
Using the data listed in Table \ref{tb1} and adopting a solar motion ($U$,  $V$, $W$)$_{\sun}=$ (10.0, 5.2, 7.2) km s$^{-1}$ \citep{de98}, the local standard of rest (LSR) velocities of PNs were determined. The LSR velocities were then corrected to the Galactic standard of rest (GSR) by adopting the galactocentric distance of the Sun $R_{\sun}=8.0$ kpc and a rotation velocity of the LSR at the position of the Sun of 220 km s$^{-1}$ \citep[and reference therein]{wu09}. The observed GSR positions and velocities of the eight nearby PNs are listed in Table \ref{tb2}. The position coordinates $x$, $y$, $z$ and velocity components $U$, $V$, $W$ refer to a galactocentric righthanded Cartesian coordinate system with the directions pointing to the Galactic center, direction of rotation, and the Galactic north pole. The errors in the velocity components listed in Table \ref{tb2} include the errors in distances, proper motions, and radial velocities of PNs listed in Table \ref{tb1}. The LSR velocities of the PNs are listed in the last column.

We employed the axisymmetric Galactic gravitational potential models of \citet[hereafter AS91]{as91} and \citet[hereafter FSC96]{fsc96}. These two models consist of three components: bulge, disk, and spherical halo. Both models admit two conserved quantities: the total energy $E$ and the $z$-component $J_{z}$ of the angular momentum vector. The disks are very different between the two models \citep{wu09}. Using the initial conditions listed in Table \ref{tb2}, the orbits of the PNs were calculated backward in a time interval of 2 Gyr with the two given gravitational potential models. The Bulirsch-Stoer algorithm of \citet{pr92} was used for the integration. The relative change in the total energy over the 2 Gyr integration time is of the order of $10^{-14}$.

The orbital parameters $e$, $z_{\textrm{max}}$, and $T_{z}$ calculated with the AS91 and FSC96 models are listed in Table \ref{tb3}. $e=(R_{\textrm{a}}-R_{\textrm{p}})/(R_{\textrm{a}}+R_{\textrm{p}})$ is the orbital eccentricity, where $R_{\textrm{a}}$ and $R_{\textrm{p}}$ are apogalactic and perigalactic distances from the Galactic center, respectively, which are determined from the average maximum and minimum galactocentric distances in the calculated orbits within the integration time of 2 Gyr. $z_{\textrm{max}}$ is the average maximum vertical distance above the Galactic plane \citep{wu09}. $T_{z}$ is the mean time interval for the PN to cross the Galactic plane from one $z_{max}$ to the other in the opposite direction \citep{wu09}. The $z$-component $J_{z}$ of the angular momentum vector of each PNs is also listed in Table \ref{tb3}. For all eight PNs in our sample, Tables \ref{tb21} and \ref{tb3} indicate that the mean time $T_{z}$ for each PN is much larger than its kinematic age.

The errors of orbital parameters listed in Table \ref{tb3} were calculated based on a Monte Carlo method. Because the intrinsic uncertainties of the orbits within the 2 Gyr integration interval are very small, only the errors in the input data were used to calculate the errors of the derived orbital parameters \citep{wu09}. For each PN, the initial condition was generated by adding Gaussian deviates to the observed proper motion, radial velocity, and distance. The errors for the input data listed in Table \ref{tb2} are adopted as the standard deviations. For each PN, the errors for its orbital parameters were calculated based on 1000 separate integrations.

The average difference for the orbital parameter $e$, $z_{\textrm{max}}$, and $T_{z}$ derived from the AS91 and FSC96 models are $0.006\pm0.003$, $-54.1\pm13.2$ pc, and $-10.5\pm0.3$ Myr, respectively. Table \ref{tb3} shows that, the relative differences in the derived orbital parameters due to the adopted different potential models are smaller than those derived from observational errors. The relative large differences in $z_{\textrm{max}}$ and $T_{z}$ for the two models are due to the fact that the mass of the disk in the FSC96 model is smaller than that in the AS91 model \citep{wu09}.

\begin{figure*}
\begin{center}
\includegraphics[width=140mm,height=190mm]{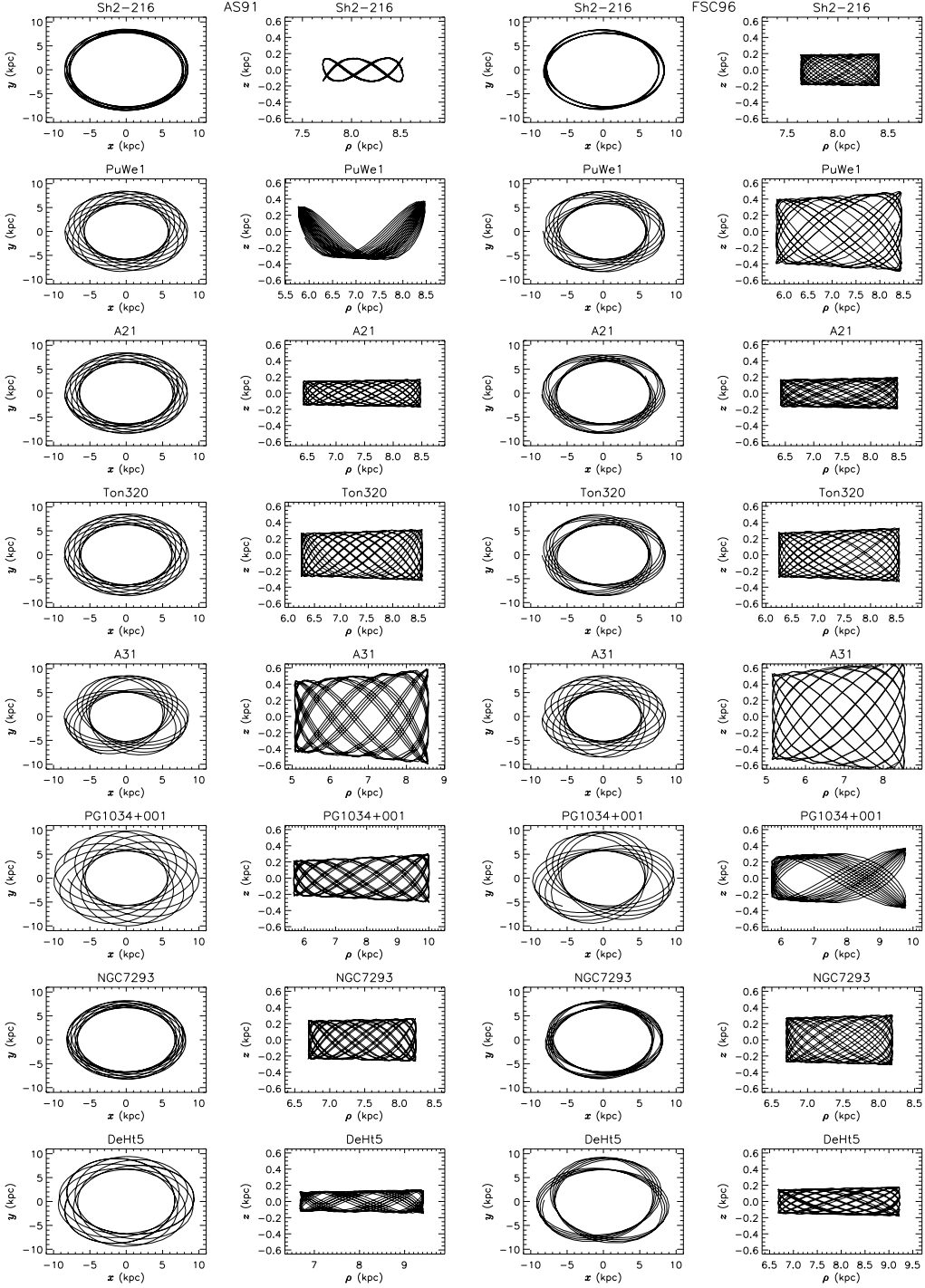}
\caption{Meridional Galactic orbits and orbits projected on to the Galactic plane in the time interval of 2 Gyr for the eight PNs in our sample calculated with the AS91 and FSC96 models.\label{fg1}}
\end{center}
\end{figure*}

The orbits calculated with the AS91 and FSC96 models in the time interval of 2 Gyr are presented in Figure \ref{fg1}, where $\rho=\sqrt{x^2+y^2}$. For all PNs in our sample, their orbits projected on the Galactic plane in Figure \ref{fg1} clearly indicate the periodic motions of PNs in the plane. For most of the PNs, the meridional orbits in Figure \ref{fg1} are of a boxy-like type and PNs move in the meridional plane within the limited regions almost filling the boxes. However, the meridional orbits of Sh 2-216 and PuWe 1 in the AS91 model and that of PG1034+001 in the FSC96 model indicate that they only move with special paths within the meridional plane. Thus, the weird orbits of Sh 2-216 and PuWe 1 derived from the AS91 model and that of PG1034+001 derived from the FSC96 model are not used in the following analysis. These weird orbits are caused by the Galactic models adopted. Thus, we need to check each model for every object in detail, which is  beyond the scope of this paper. This problem is under investigation and will be discussed in a subsequent paper.

\citet{sc10} re-examined the velocity of the Sun with respect to the LSR. They found that the classical determination of its component $V_{\sun}$ in the direction of Galactic rotation is undermined by the metallicity gradient in the disk. Using a chemodynamical model which accounts for these effects, they obtained ($U$,  $V$, $W$)$_{\sun}=$ (11.1, 12.24, 7.25) km s$^{-1}$. In particular, $V_{\sun}$ is 7 km s$^{-1}$ larger than that of \citet{de98} adopted in this study. Using the new parameters of the Sun derived by \citet{sc10}, we re-calculated the orbits of the eight PNs in our sample with the AS91 and FSC96 models. Using the AS91 model, the mean differences for the orbital parameters $e$, $z_{\textrm{max}}$, and $T_{z}$ derived with the different parameters of the Sun \citep{sc10,de98} are $-0.018\pm0.006$, $3.9\pm0.7$ pc, and $0.9\pm0.1$ Myr, respectively. The corresponding mean differences for $e$, $z_{\textrm{max}}$, and $T_{z}$ with the FSC96 model are $-0.018\pm0.005$, $2.0\pm0.9$ pc, and $1.5\pm0.2$ Myr, respectively. The effects of adopting different parameters of the Sun are very small for the derived orbital parameters with different Galactic models.

\subsection{Population membership}
\citet{wu09} analyzed the orbital eccentricity distributions for different Galactic populations and found that no thin-disk star has an eccentricity greater than 0.3 and no thick-disk star has an eccentricity less than 0.1. The orbital eccentricities of the thin-disk and the thick-disk stars overlap between 0.1 and 0.3. The mean $e$ for the thin-disk stars is $\sim 0.1$ and the mean for thick-disk stars is $\sim 0.4$. In our sample, only Sh 2-216 has $e$ less than 0.1 which indicates it is a thin-disk PN. It is difficult to assign the population membership for the other seven PNs based only on their orbital parameter $e$.

\begin{figure*}
\begin{center}
 \includegraphics[width=140mm,height=90mm]{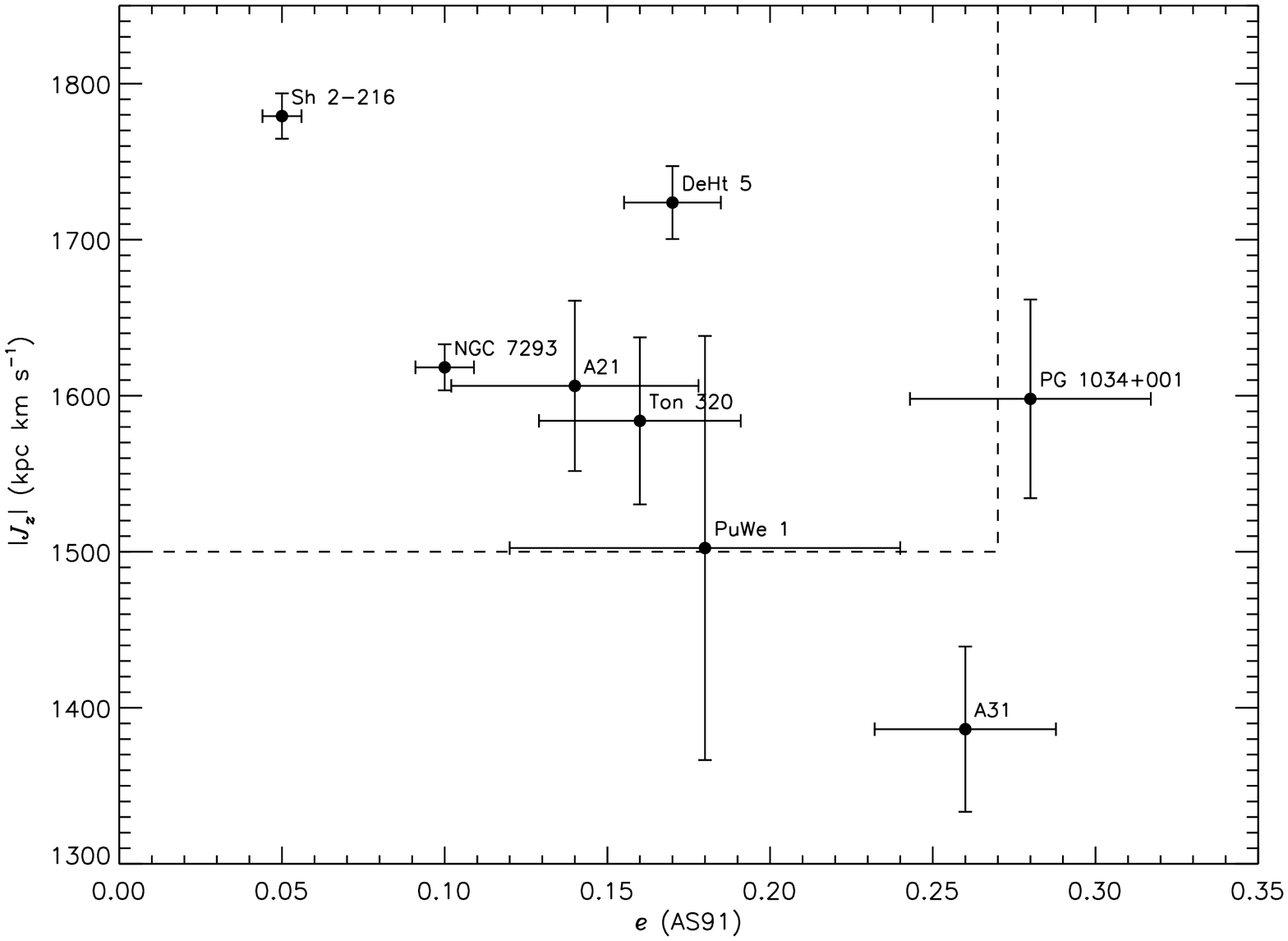}
\caption{$|J_{z}|$ --- $e$ diagram of the eight PNs in our sample. The top left region enclosed by two dashed lines is occupied by the thin-disk population.\label{fg2}}
\end{center}
\end{figure*}

\citet{pa03,pa06} presented kinematic criteria for distinguishing thin-disk, thick-disk, and halo populations in their DA white dwarf sample. Their kinematic criteria were deduced from a sample of F and G stars chosen from the literature, for which chemical criteria can be used to distinguish different populations. Their kinematic population classification scheme is based on the position in the $U-V$ velocity diagram, the position in the $J_{z}-e$ diagram, and the Galactic orbits themselves. Among these three kinematic criteria, the $J_{z}-e$ diagram can distinguish different populations more clearly (comparing Figures 2 and 3 of \citep{pa06}). In the $J_{z}-e$ diagram of \citet{pa06}, the thin-disk stars lie in a V-shaped area with $e<0.27$ and $J_{z}$ around 1800 kpc km s$^{-1}$. The thick-disk stars possess $e>0.27$ and $J_{z}<1500$ kpc km s$^{-1}$. Because of our adopted right-handed orientation of the coordinate system, the direction of $J_{z}$ vector is opposite to that of \citet{pa03,pa06}. We present the $|J_{z}|-e$ diagram in Figure \ref{fg2}, the adopted orbital parameters $e$ are derived from the AS91 model. Because of the weird orbits of Sh 2-216 and PuWe 1 in the AS91 model, their parameters $e$ derived from the FSC96 model are adopted in Figure \ref{fg2}. The error of $e$ for each PN in Figure \ref{fg2} includes the error due to the uncertainty in the input data and that derived with different Galactic models. Due to the weird orbits of Sh 2-216, PuWe 1, and PG1034+001 in different models, their errors of $e$ in different models cannot be confirmed. Thus, only errors derived from the input data are considered for the three above-mentioned PNs.

In Figure \ref{fg2}, dashed lines divide this figure into two regions for different populations. The top left region enclosed by these two dashed lines is the one for the thin-disk population. The other region is occupied by the thick-disk population. Figure \ref{fg2} shows that PG 1034+001 and A31 lie in the region of the thick-disk population and Sh 2-216, DeHt 5, NGC 7293, A21, and Ton 320 belong to the thin-disk population. PuWe 1 has low $|J_{z}|$ and $e\sim 0.2$ which indicates it likely belongs to the thick-disk population, but the large errors in its orbital parameters make its population classification more uncertain.

\section{Comments on individual objects}
\subsection{Sh 2-216}
The orbit of Sh 2-216 was first calculated by \citet{ke04} with the AS91 model. Using the method of \citet{pa03}, \citet{ke04} identified Sh 2-216 as a thin-disk PN. They used the trigonometrical distance derived by \citet{ha97} which is consistent with that derived by \citet{ha07}. The proper motions ($\mu_{\alpha}$, $\mu_{\delta}\cos\alpha$) $=$ (21.7, $-12.7$) mas yr$^{-1}$ used by \citet{ke04} were taken from the UCAC 2 catalog which are also consistent with those taken from the UCAC 3 catalog. \citet{ke04} used the radial velocity of 11.9 km s $^{-1}$ derived by \citet{tw92} which is half of that derived by \citet{ra07}. \citet{ra07} have pointed out that the result of \citet{tw92} may be less accurate because the CS of this PN was possibly not well centered in the observation aperture. Using the radial velocity of \citet{tw92}, we re-calculated the orbit of Sh 2-216 in the AS91 model, the orbital parameters $e$, $z_{\textrm{max}}$, and $T_{z}$ are derived as 0.03, 142 pc, and 33.3 Myr, respectively. The adopted different radial velocities of Sh 2-216 do not affect the derived orbital parameters with the same Galactic model. Using the orbital parameters derived with the radial velocity of \citet{tw92}, based on its position in Figure \ref{fg2}, Sh 2-216 is still identified as a thin-disk PN, which is also consistent with that of \citet{ke04}. Therefore, using the radial velocity of \citet{tw92} will not affect the population membership determination of Sh 2-216.

The interaction between Sh 2-216 and the ISM was studied by \citet{tw95} in detail. A region of enhanced emission in the east of the PN and the displacement of the CS with respect to the center of the PN clearly indicate the PN-ISM interaction. The P.A. of the proper motions of Sh 2-216 also indicates that the direction of the movement of the PN is consistent with the position of the PN-ISM interaction region. Using the kinematic age 0.45 Myr and the $W$ velocity of Sh 2-216 listed in Table \ref{tb2}, this PN only moved $6.2$ pc away from its birthplace in the vertical direction. We have pointed out in the previous section that the kinematic age of this PN is much smaller than the mean time $T_{z}$ for it to cross the Galactic plane.

 Assuming that the ISM in the vicinity of the Sun has the same rotation velocity of the LSR, the LSR velocities of the PNs listed in Table \ref{tb2} can be considered as the approximately relative velocities of those PNs with respect to their nearby ISM. Using a relative velocity of $\sim$ 20 km s$^{-1}$,  with an estimated electron density of $n_{\textrm{PN}}\sim$ 8 cm$^{-3}$ \citep{ra08} and an expansion velocity of $\sim$ 4 km s$^{-1}$ \citep{re85}, the models of BSS and SBS should need a local ISM density of $n_{\textrm{ISM}}\sim$ 1.4 cm$^{-3}$ to explain the observed PN-ISM interaction of Sh 2-216. According to the model of \citet{ka98}, the density of the diffuse ISM at the position of Sh 2-216 is 0.12 cm$^{-3}$ is much lower than that estimated from the models of BSS and SBS. \citet{tw95} estimated a relative velocity of only $\sim$ 5 km s$^{-1}$; an unusually high ISM density of $n_{\textrm{ISM}}\sim$ 4 cm$^{-3}$ can be estimated with this much lower relative velocity. If we use the radial velocity of \citet{tw92}, based on the models of BSS and SBS, the local ISM density can be derived as 2.2 cm$^{-3}$, which is still much larger than that derived from the model of \citet{ka98}.

Based on the above analysis, we find that the local density of ISM near Sh 2-216 derived from the models of BSS and SBS is much larger than that derived from the model of \citet{ka98}. The model of \citet{ka98} represents the large-scale ISM distribution in the Galaxy, the values listed in Table \ref{tb21} are the average and smoothed density of ISM at the position of Sh 2-216. In fact, large density perturbations on small scales of ISM are observed \citep{di90}, hence the model of \citet{ka98} may underestimate the local density of ISM interacting with Sh 2-216. On the other hand, if the density of ISM derived from the model of \citet{ka98} is correct at the position of Sh 2-216, other factors must be considered in the models of BSS and SBS to explain the observed PN-ISM interaction for Sh 2-216. \citet{tw95} pointed out that the ISM magnetic field is needed to mold the interaction between Sh 2-216 and its nearby ISM.

 \citet{ra08} presented 1420 MHz polarization images of Sh 2-216. An arc of low polarized intensity in the northeast portion was observed and was coincident with the optically identified PN-ISM interaction region. Several polarization-angle ``knots'' appeared along the arc. \citet{ra08} estimated the rotation measure through the knots and derived a line-of-sight magnetic field in the interaction region as $5.0\pm2.0$ $\mu$G. This result is consistent with that estimated by \citet{tw95}.

In the simulations of \citet{vi03} and WZO, with a low relative velocity $\sim$ 20 km s$^{-1}$, the PN-ISM interaction can be observed for a lower ISM density of $n_{\textrm{ISM}}=0.01$ cm$^{-3}$. The magnetic field is also not considered in their simulations. The CS has been displaced from the center of the PN, so Sh 2-216 is in the WZO 3 stage.

\subsection{PuWe 1}
Figure 2 of \citet{tw96} shows that PuWe 1 has a spherical enhanced emission shell interacting with the ISM, but the northwest sector of the shell  is  dimmer than the other portions. The P.A. of the proper motions of PuWe 1 indicates that this PN is moving to the southeast which is consistent with the direction of the PN-ISM interaction region.  The orbit of PuWe 1 shows that it is moving toward its maximum distance from the plane of the Galaxy. Using the kinematic age 0.04 Myr and the $W$ velocity of PuWe 1 listed in Table \ref{tb2}, this PN only moved $1.1$ pc away from its birthplace in the vertical direction.

The relative velocity of PuWe 1 is $\sim$ 50 km s$^{-1}$ which is the typical velocity in the models of BSS and SBS for disk PNs. Using the average volume filling factor of Galactic disk PNs $\varepsilon=0.37$ \citep{bo94}, the mean electron density of PuWe 1 can be estimated as 3.1 cm $^{-3}$ \citep{ph98}. According to BSS and SBS, with an expansion velocity of 27 km s$^{-1}$, the ISM density can be calculated as $\sim 0.05$ cm $^{-3}$ which is close to the average diffuse ISM density $n_{\textrm{ISM}}=0.1$ cm $^{-3}$ derived from the model of \citet{ka98} at the position of PuWe 1. Because the CS is still in the center of the PN, PuWe 1 is in the WZO 2 stage.

\subsection{A21}
Both the [\ion{O}{3}] \citep{kw83} and H$\alpha$ \citep{hu99} images of A21 show that this PN is a one-side nebula, whose bright filaments are in the southeast and a faint rim is on the opposite side. The bright filamentary structure in the southeast indicates the PN-ISM interaction. The P.A. of the proper motions of this PN is consistent with the direction of the PN-ISM interaction regions. \citet{tw96} explained the one-side structure of A21 as the result of a highly inhomogeneous ISM in the vicinity of this PN.

The orbits of A21 indicate that it is a typical thin-disk object. Using the derived kinematic age 0.02 Myr and a $W$ velocity of A21, this PN only moved $0.2$ pc away from its birthplace in the vertical direction. Different orbital parameters derived with different Galactic models cannot affect the PN-ISM interaction of A21.

The electron densities of A21 in six regions of the bright filaments were derived by \citet{kw83} and the mean is $206\pm42$ cm$^{-3}$. Using a relative velocity of $\sim 30$ km s$^{-1}$ and an expansion velocity of $\sim 30$ km s$^{-1}$, according to BSS and SBS, the density of ISM in the vicinity of A21 can be estimated as $\sim 5$ cm$^{-3}$. According to the model of \citet{ka98}, the density of ISM at the position is 0.09 cm $^{-3}$, which is much smaller than that derived from the models of BSS and SBS. BSS has pointed out that an asymmetric \ion{H}{2} region surrounding A21 should exist, which would explain the observed shape of the PN and the estimated high ISM density. The model of \citet{ka98} may underestimate the local density of ISM in the vicinity of A21. Because the CS of A21 has been displaced from the center, this PN is in the WZO 3 stage.

\subsection{Ton 320}
This PN was found during a search around the hot H-rich white dwarf Ton 320 \citep{tw94}. With a distance of 532 pc derived by \citet{ha07}, this PN has a linear diameter of 4.6 pc, which is bigger than Sh 2-216. The [\ion{N}{2}] image of Ton 320 \citep{tw94} shows that this PN is very faint and it has a similar one-side filamentary structure as that of A21. The bright filaments in the southeast indicate it is interacting with the ISM, no faint filaments can be observed in the opposite direction in the same [\ion{N}{2}] image. The P.A. of the proper motions of Ton 320 is consistent with the direction of the PN-ISM interaction region.

Using the kinematic age 0.11 Myr and the $W$ velocity of Ton 320, this PN only moved $1.1$ pc away from its birthplace in the vertical direction. Because there is no electron density data available for Ton 320, we cannot estimate the ISM density with the models of BSS and BSB. However, using the diffuse ISM density $\sim0.1$ cm$^{-3}$ at the position of Ton 320 derived from the model of \citet{ka98}, with a relative velocity of $\sim 40$ km s$^{-1}$ and a mean expansion velocity of PNs $\sim20$ km s$^{-1}$ \citep{we89}, according to the models of BSS and SBS, the mean electron density of Ton 320 can be estimated as 3.6 cm$^{-3}$. This electron density is close to those of Sh 2-216 and PuWe 1. All three of those PNs have a large size and extremely low surface brightness, so the estimated very low electron density for Ton 320 is reasonable. The very large size and very low density of Ton 320 also indicate that it has evolved to its later stage. The CS is displacing from the center of the PN, which indicates that Ton 320 is in the WZO 3 stage.

\subsection{A31}
The H$\alpha$ and [\ion{N}{2}] images of A31 \citep{tw94} show that its southern rim is brighter and sharp-edged, whereas the northern side is faint and diffuse, which indicates its interaction with the ISM. The orbits of A31 indicate it is a thick-disk object and is moving toward the Galactic plane. Using the kinematic age 0.13 Myr and the $W$ velocity of A31, this PN only moved $3.9$ pc away from its birthplace in the vertical direction. The P.A. of the proper motions of this PN is also consistent with the direction of the PN-ISM interaction region.

Using the average filling factor of disk PNs $\varepsilon=0.37$ \citep{bo94}, the mean electron density of A31 can be estimated as 11.4 cm$^{-3}$ \citep{ph98}. Using the relative velocity of $\sim 70$ km s$^{-1}$ and expansion velocity of $\sim 10$ km s$^{-1}$, according to the models of BSS and SBS, the density of ISM in the vicinity of A31 can be estimated as 0.16 cm$^{-3}$ which is close to the average density $n_{\textrm{ISM}}\sim 0.1$ cm$^{-3}$ derived from the model of \citet{ka98} for the diffuse ISM at the position of A31.

\subsection{PG 1034+001}
The orbit of PG 1034+001 was calculated by \citet{ra04} with the AS91 Galactic model. The radial velocity and proper motions of PG 1034+001 adopted by \citet{ra04} are consistent with those adopted in this paper, but the distance of  155 pc adopted by \citet{ra04} is derived based on spectroscopic observation \citep{we95} and is less than our adopted distance of 210 pc. Using the method of \citet{pa03} and based on its position in the $U$ -- $V$ velocity diagram, $J_{z}$ -- $e$ diagram, and the Galactic orbit,  \citet{ra04} found that PG 1034+001 is a thin-disk star. Using the improved criteria of \citet{pa06} for distinguishing different populations, we find that this star is a thick-disk object. The different population membership determination is mainly due to the adopted different distances for this star.

The largest known PN surrounding the DO white dwarf PG 1034+001 with a diameter of $\sim 2\degr$ was first found by \citet{he03} based on the investigation of spectra from the Sloan Digital Sky Survey \citep[SDSS;][]{yo00}. A careful inspection of the Southern H-Alpha Sky Survey Atlas \citep[SHASSA;][]{ga01} has shown that this PN is surrounded by an elliptical emission shell with a diameter of $\sim 6\degr$, and a fragment of a wider shell with a diameter of $\sim 10\degr$ is also visible in the northeast \citep{ra04}. Using the kinematic age 0.54 Myr and the $W$ velocity of PG 1034+001 listed in Table \ref{tb2}, this PN moved $10.2$ pc away from its birthplace in the vertical direction; this distance is only half of its linear size.

\citet{ra04} noted that the major axis of the halo of the outer shell with a diameter of $\sim 6\degr$ has a P.A. of $\sim 310\degr$ which is consistent with the P.A. of the proper motions of $290\degr$. They pointed out that this structure can be understood as an advanced PN-ISM interaction, but the P.A. of the proper motions is opposite to that of the strongest emission region in the inner nebula found by \citet{he03}. \citet{ra04} also pointed out that the DO white dwarf PG 1034+001 has probably experienced a final helium-shell flash in its past, which produces  the second PN, but according to the simulations of WZO,  the outer shell with a diameter of $\sim 6 \degr$ is probably the bow shock structure formed by the interaction between the slow AGB wind and the ISM; the inner shell with a diameter of $\sim 2\degr$ is the PN shell formed by the interaction between the fast post-AGB wind and the slow AGB wind material. The PN is still within the bubble of undisturbed AGB wind material and unaffected by the ISM interaction, which indicates it is in the WZO 1 stage. Because this PN is still in the WZO 1 stage and no electron density of this PN are available, the models of BSS and BSB cannot be used to explain the PN-ISM interaction and to derive the local density of ISM for this PN.

\subsection{NGC 7293}
NGC 7293 (Helix nebula) has a bright complex helical structure which is composed of at least two bipolar nebulae, with their poles at different orientations along P.A.s $125\degr$ and $50\degr$, respectively \citep{me08}. The very deep H$\alpha+$[\ion{N}{2}] image of NGC 7293 obtained by \citet{me05} shows that there is a bow-shaped filament at the northeast of the bright helical filaments of this PN, and a faint jet-like feature at the opposite side of the PN is also visible. \citet{ke04} calculated the orbit of NGC 7293 with the AS91 model and found that this PN is a thin-disk object. In this paper, we also identified NGC 7293 to be a thin-disk PN, which is consistent with the result of \citet{ke04}. Using the kinematic age 0.03 Myr and the $W$ velocity of PG NGC 7293 listed in Table \ref{tb2}, this PN moved $0.4$ pc away from its birthplace in the vertical direction.

\citet{tw96} noted that the main shell of NGC 7293 does not show any signs of the PN-ISM interaction. They also pointed out that the asymmetric halo of this PN is due to the symmetrically ejected material from the CS which runs into the higher density northeast region of its southwest counterpart. With the average electron density of this PN $\sim 630$ cm$^{-3}$ \citep{bo94}, a relative velocity of $\sim 30$ km s$^{-1}$, and an expansion velocity of $\sim 20$ km s$^{-1}$, according to the models of BSS and SBS, the density of ISM in the vicinity of NGC 7293 can be estimated as $\sim 22$ cm$^{-3}$ which is much larger than $n_{\textrm{ISM}}=0.08$ derived from the model of \citet{ka98} for the diffuse ISM.

According to the simulations of WZO, the bow-shaped structure in the halo is probably formed by the interaction between the AGB wind and the ISM. The P.A. of the proper motions of the PN is consistent with the outer bow-shaped region, which indicates a probable wind-ISM interaction. The inner bright helical shell is not affected by the ISM interaction, and this PN is in the WZO 1 stage.

\subsection{DeHt 5}
The H$\alpha$ and [\ion{N}{2}] images of DeHt 5 \citep{tw96} show a large amount of diffuse material surrounding the bright filaments of this PN, which indicates that the PN-ISM interaction has destroyed the intrinsic morphology of this PN. The very low expansion velocity of 5 km s$^{-1}$ also indicates that DeHt 5 has evolved into its later stage. Using the kinematic age of 0.09 Myr and the $W$ velocity of DeHt 5 listed in Table \ref{tb2}, this PN moved $1.0$ pc away from its birthplace in the vertical direction.

Using the relative velocity of $\sim 60$ km s$^{-1}$ and the expansion velocity of 5 km s$^{-1}$, according to the models of BSS and SBS, and using the diffuse ISM density $\sim0.1$ cm$^{-3}$ derived from the model of \citet{ka98}, the mean electron density of DeHt 5 can be estimated as 4.5 cm$^{-3}$ which is close to those of Sh 2-216, PuWe 1, and Ton 320. This very low density also indicates that DeHt 5 has evolved into its later stage.

\section{Discussion and conclusions}
The orbits of eight nearby PNs are calculated with the AS91 and FSC96 Galactic models. The small number of PNs in the present sample is limited by the available observation data for PNs in the Galaxy. In our present sample, only PNs with direct parallax distances are considered. The errors of the derived orbital parameters are calculated with a Monte Carlo method. We also compared the orbital parameters derived with different Galactic models. We find that the uncertainty of the derived orbital parameters is mainly from the observational errors in the input data. 

According to the kinematic criteria of \citet{pa03,pa06} for distinguishing different populations in the Galaxy and using the $J_{z}-e$ diagram, we find that Sh 2-216, DeHt 5, NGC 7293, A21, and Ton 320 belong to the thin-disk population and PG 1034+001 and A31 belong to the thick-disk population. PuWe 1 probably belongs to the thick-disk population, but with the large errors of its derived orbital parameters, its population classification becomes very uncertain. The present sample of PNs is very small, hence we cannot come to any conclusion about the population distribution of Galactic PNs in the disk. Population classification for the eight PNs in our sample indicates that most of the PNs formed are closer to the Galactic plane. The dramatic enlargement of the sample of PNs with reliable parallaxes will only be expected when the data of the \textit{Gaia} mission are available. The primary objective of the \textit{Gaia} mission is to survey more than one billion stars between 6 and 20 mag. The expected accuracies for parallaxes, proper motions and positions are 7 -- 25 $\mu$as down to 15 mag and a few hundred $\mu$as at 20 mag \citep{li10}. The magnitudes of 96\% CS in the catalog of \citet{ke03} are less than 20.0; the precise distances and proper motions for those stars obtained with the \textit{Gaia} mission can be used to study their kinematics and dynamics in the Galaxy.

We analyzed the interaction between the PNs in our sample and their nearby ISM. The PN-SIM interaction are observed in the eight PNs in our present sample. The P.A.s of the proper motions of the PNs are also consistent with the directions of the PN-ISM interaction regions. For all eight of the PNs in our sample, their kinematic ages are much smaller than the time $T_{z}$ for them to cross the Galactic plane, and the distances that they moved from their birthplaces are all smaller than their linear size. Different orbits derived with different Galactic models cannot affect the PN-ISM interaction for each PN.

There are two main models to explain the PN-ISM interaction. One is presented by BSS and SBS. In their models, the PN  interacts with the ISM directly, and the PN-ISM interaction can be observed only when the density of the PN drops below some critical value.  The magnetic field is also important in the models of BSS and SBS. The other one is the “triple-wind" model presented by WZO. In their model, the slow AGB wind formed in the early evolution phase of the AGB star interacts with the ISM, and then the PN interacts with the bow shock formed during the AGB wind-ISM interaction. So, in the model of WZO, the PN-ISM interaction can be observed in the early evolution phase of the PN. The magnetic field is not considered in the model of WZO. For both models, the relative velocity of the PN and the density of the ISM in the vicinity of the PN are very important parameters for studying the PN-ISM interaction.

We use the above-mentioned models to explain the observed PN-ISM interaction in the eight PNs in our sample. Using the models of BSS and SBS, the local densities of ISM in the vicinity of PuWe 1 and A31 are derived and are consistent with those derived from the model of \citet{ka98}. For Ton 320 and DeHt 5, no electron density data are available. Using the local densities of ISM for these two PNs derived from the model of \citet{ka98}, the electron densities of these two PNs are derived from the models of BSS and SBS. Compared with the observed electron densities of other PNs which have evolved into their later stages, the derived electron densities of these two PNs are reasonable. For the above four PNs, the models of BSS and SBS yield consistent results from the model of \citet{ka98}, but for Sh 2-216 and A21, the derived local densities of ISM in the vicinity of these two PNs with the models of BSS and SBS are larger than those derived from the model of \cite{ka98}. For Sh 2-216, considering a magnetic field in the position of this PN in the models of BSS and SBS can explain the observed PN-ISM interaction. For A21, an asymmetric \ion{H}{2} region surrounding this PN can explain the predicted high ISM density. For these two PNs, the model of \citet{ka98} will underestimate the local densities of ISM. The model of \citet{ka98} represents the large-scale and average ISM distribution in the Galaxy; it cannot predict the local density of ISM very precisely. Therefore, with the models of BSS and SBS, the local ISM densities in the vicinity of evolved PNs can be derived with the observed PN-ISM interaction \citep{bo90}. Because PG 1034+001 and NGC 7293 are still in their early stage of evolution, the models of BSS and SBS can not used to derive the ISM densities for these two PNs. On the other hand, according to the model of WZO, Sh 2-216, A21, and Ton 320 are in the WZO 3 stage, PG 1034+001 and NGC 7293 are in the WZO 1 stage, and PuWe 1 is in the WZO 2 stage.

\begin{deluxetable*}{lllllccccc}
\tablewidth{0pt}
\tablecaption{The observed data of eight nearby PNs in our sample.\label{tb1}}
\tablehead{\colhead{Name}&\colhead{$\alpha$ (2000.0)}&\colhead{$\delta$ (2000.0)}&\colhead{$V$}&\colhead{$\pi_{\textrm{abs}}$}&\colhead{$\mu_{\alpha}\cos\delta$}&\colhead{$\mu_{\delta}$}&\colhead{P.A.}&\colhead{$v_{\textrm{rad}}$}&\colhead{Reference}\\ \colhead{}&\colhead{(hh : mm : ss)}&\colhead{(dd : mm : ss)}&\colhead{(mag)}&\colhead{(mas)}&\colhead{(mas yr$^{-1}$)}&\colhead{(mas yr$^{-1}$)}&\colhead{(\degr)}&\colhead{(km s$^{-1}$)}}
\startdata
Sh 2-216   &04 : 43 : 21.3&$+46$ : 42 : 05.8&12.630&$7.76\pm0.33$                 &$22.9\pm1.8$ &$-11.6\pm3.5$\tablenotemark{b}&116&$20.6\pm1.5$&$7$\\
PuWe 1     &06 : 19 : 34.3&$+55$ : 36 : 42.3&15.534&$2.74\pm0.31$                 &$19.6\pm14.0$&$-24.8\pm8.0$&141&$18.3\pm1.2$&$3$\\
A21        &07 : 29 : 02.7&$+13$ : 14 : 48.8&15.962&$1.85\pm0.51$                 &$2.4\pm2.0$  &$-10.0\pm2.0$&166&$28.8\pm5.2$&$2$\\
Ton 320    &08 : 27 : 05.6&$+31$ : 30 : 08.5&15.702&$1.88\pm0.33$                 &$-4.5\pm2.0$ &$-13.6\pm2.0$&198&$33.8\pm3.5$&$4$\\
A31        &08 : 54 : 13.2&$+08$ : 53 : 53.0&15.519&$1.61\pm0.21$\tablenotemark{a}&$-18.4\pm2.0$&$-16.5\pm2.0$&228&$44.0\pm1.0$&$5$\\
PG 1034+001&10 : 37 : 03.8&$-00$ : 08 : 19.2&13.211&$4.75\pm0.53$                &$-82.4\pm10.1$&$31.2\pm10.0$\tablenotemark{b}&290&$50.8\pm1.3$&$6$\\
NGC 7293   &22 : 29 : 38.5&$-20$ : 50 : 13.7&13.525&$4.64\pm0.27$\tablenotemark{a}&$27.4\pm1.4$ &$-5.4\pm1.5$\tablenotemark{b}&101&$-28.2\pm3.0$&$2$\\
DeHt 5     &22 : 19 : 33.7&$+70$ : 56 : 03.3&15.474&$2.90\pm0.15$\tablenotemark{a}&$-17.4\pm2.0$&$-19.8\pm2.0$&221&$-35.4\pm2.5$&$1$\\
\enddata
\tablenotetext{a}{Taken from Table 5 of  \citet{be09}.}
\tablenotetext{b}{Taken from the UCAC 3 catalog.}
\tablerefs{(1) \citet{ba01}; (2) \citet{du98}; (3) \citet{gi86}; (4) \citet{go05}; (5) \citet{hi90}; (6) \citet{ho98}; (7) \citet{ra07}}
\end{deluxetable*}

\begin{deluxetable*}{lccccccccc}
\tablewidth{0pt}
\tablecaption{The sizes, electron densities, and expansion velocities of eight nearby PNs in our sample and the number densities of ISM at the positions of these PNs.\label{tb21}}
\tablehead{\colhead{Name}&\colhead{angular diameter}&\colhead{linear diameter}&\colhead{distance}&\colhead{$n_{\textrm{PN}}$}&\colhead{$n_{\textrm{ISM}}$ (WNM)}&\colhead{$n_{\textrm{ISM}}$ (WIM)}&\colhead{$v_{\textrm{exp}}$}&\colhead{kinematic age}&\colhead{Reference}\\ \colhead{}&\colhead{(arcmin)}&\colhead{(pc)}&\colhead{(pc)}&\colhead{(cm$^{-3}$)}&\colhead{(cm$^{-3}$)}&\colhead{(cm$^{-3}$)}&\colhead{(km s$^{-1}$)}&\colhead{Myr}&\colhead{}}
\startdata
Sh 2-216   &100&3.7&128.9&8.0&0.10&0.02&4&0.45&5\\
PuWe 1     &20&2.1&365.0&3.1&0.08&0.02&27&0.04&1\\
A21        &9&1.4&540.5&206.0&0.07&0.02&32&0.02&4\\
Ton 320    &30&4.6&531.9&&0.05&0.02&&0.11\tablenotemark{b}&\\
A31        &16&2.9&621.1&11.4&0.04&0.02&11&0.13&2\\
PG 1034+001&360\tablenotemark{a}&22.0&210.5&&0.07&0.02&&0.54\tablenotemark{b}&\\
NGC 7293   &18&1.1&215.5&630.0&0.06&0.02&21&0.03&3\\
DeHt 5     &9&0.9&344.8&&0.08&0.02&5&0.09&1\\
\enddata
\tablenotetext{a}{Taken from \citet{ra04}.}
\tablenotetext{b}{Use the mean expansion velocity of Galactic PNs 20 km s$^{-1}$ \citep{we89}.}
\tablerefs{(1) \citet{gi86}; (2) \citet{hi90}; (3) \citet{me08}; (4) \citet{re81}; (5) \citet{re85}}
\end{deluxetable*}

\begin{deluxetable*}{llllccccc}
\tablewidth{0pt}
\tablecaption{The present positions and velocities of eight nearby PNs in the Galaxy.\label{tb2}}
\tablehead{\colhead{Name}&\colhead{$x$}&\colhead{$y$}&\colhead{$z$}&\colhead{$U$}&\colhead{$V$}&\colhead{$W$}&\colhead{$v_{\textrm{LSR}}$}\\ \colhead{}&\colhead{(pc)}&\colhead{(pc)}&\colhead{(pc)}&\colhead{(km s$^{-1}$)}&\colhead{(km s$^{-1}$)}&\colhead{(km s$^{-1}$)}&\colhead{(km s$^{-1}$)}}
\startdata
Sh 2-216   &$-8119.9\pm  5.1$&$   47.3\pm  2.0$&$    1.1\pm  0.0$&$ -14.4\pm 1.6$&$ 219.2\pm 1.8$&$  13.3\pm 1.6$&$  19.6\pm 2.4$\\
PuWe 1     &$-8324.2\pm 36.7$&$  124.9\pm 14.1$&$  111.9\pm 12.7$&$ -20.8\pm 7.1$&$ 180.8\pm16.8$&$  27.3\pm22.0$&$  52.1\pm13.4$\\
A21        &$-8474.3\pm130.7$&$ -222.6\pm 61.4$&$  133.0\pm 36.7$&$  -5.5\pm 5.8$&$ 189.4\pm 8.4$&$   9.2\pm 5.3$&$  32.4\pm 3.3$\\
Ton 320    &$-8436.9\pm 76.7$&$  -88.1\pm 15.5$&$  290.3\pm 51.0$&$ -22.0\pm 4.1$&$ 187.5\pm 7.5$&$   9.6\pm 5.4$&$  40.4\pm 3.5$\\
A31        &$-8411.7\pm 53.7$&$ -335.0\pm 43.7$&$  322.6\pm 42.1$&$ -35.2\pm 4.9$&$ 163.4\pm 7.0$&$ -29.9\pm 9.3$&$  73.1\pm 5.7$\\
PG 1034+001&$-8054.0\pm  6.0$&$ -130.8\pm 14.6$&$  155.8\pm 17.4$&$ -86.9\pm13.5$&$ 197.0\pm 7.9$&$  18.6\pm 7.4$&$  91.8\pm12.9$\\
NGC 7293   &$-7905.5\pm  5.5$&$   69.0\pm  4.0$&$ -181.0\pm 10.5$&$ -23.6\pm 2.2$&$ 204.9\pm 1.9$&$  15.5\pm 2.8$&$  32.0\pm 2.1$\\
DeHt 5     &$-8121.5\pm  6.3$&$  315.1\pm 16.3$&$   69.6\pm  3.6$&$  60.6\pm 3.7$&$ 209.9\pm 2.8$&$ -10.7\pm 3.3$&$  62.4\pm 3.7$\\
\enddata
\end{deluxetable*}

\begin{deluxetable*}{llllllll}
\tablewidth{0pt}
\tablecaption{The Galactic orbital parameters of eight nearby PNs in our sample calculated with the AS91 and FSC96 models.\label{tb3}}
\tablehead{\colhead{Name}&\colhead{$J_{z}$}&\multicolumn{2}{c}{$e$}&\multicolumn{2}{c}{$z_{max}$(pc)}&\multicolumn{2}{c}{$T_{z}$(Myr)}\\
\colhead{}&\colhead{(kpc km s$^{-1}$)}&\colhead{(AS91)}&\colhead{(FSC96)}&\colhead{(AS91)}&\colhead{(FSC96)}&\colhead{(AS91)}&\colhead{(FSC96)}}
\startdata
Sh 2-216          &$ -1779.2\pm 14.5$&$0.05\pm0.01$&$0.05\pm0.01$&$ 143.0\pm 20.3$&$ 190.0\pm 27.1$&33.9&44.4\\
PuWe 1            &$ -1502.4\pm139.0$&$0.19\pm0.07$&$0.18\pm0.06$&$ 335.0\pm321.3$&$ 439.0\pm441.1$&35.1&46.5\\
A21               &$ -1606.3\pm 55.2$&$0.14\pm0.04$&$0.14\pm0.04$&$ 158.0\pm 43.7$&$ 180.0\pm 58.2$&31.7&41.7\\
Ton 320           &$ -1583.9\pm 54.1$&$0.16\pm0.03$&$0.16\pm0.03$&$ 287.0\pm 38.5$&$ 299.0\pm 44.0$&35.1&44.4\\
A31               &$ -1386.3\pm 51.5$&$0.26\pm0.03$&$0.25\pm0.02$&$ 504.0\pm154.6$&$ 620.0\pm212.6$&38.5&50.0\\
PG 1034+001       &$ -1598.0\pm 64.2$&$0.28\pm0.04$&$0.26\pm0.03$&$ 254.0\pm 72.5$&$ 311.0\pm101.5$&35.7&46.5\\
NGC 7293          &$ -1618.2\pm 15.2$&$0.10\pm0.01$&$0.10\pm0.01$&$ 248.0\pm 22.7$&$ 290.0\pm 33.5$&34.5&44.4\\
DeHt 5            &$ -1723.8\pm 23.9$&$0.17\pm0.01$&$0.16\pm0.01$&$ 130.0\pm 31.2$&$ 163.0\pm 43.6$&33.9&44.4\\
\enddata
\end{deluxetable*}

\acknowledgments
The authors thank the referee for his/her comments. This work has been supported in part by the National Natural Science Foundation of China, Nos. 10633020, 10778720, 10873016, and 10803007 and by the National Basic Research Program of China (973 Program), No. 2007CB815403. Z.-Y. W. is supported by the Young Researcher Grant of the National Astronomical Observatories, Chinese Academy of Sciences. This research has made use of the SIMBAD database and VizieR catalog access tool, operated at CDS, Strasbourg, France.

\end{document}